\documentclass{jetpl}
\twocolumn
\title{Osmotic pressure of matter and vacuum energy}

\rtitle{Osmotic pressure of matter and vacuum energy}

\sodtitle{Osmotic pressure of matter and vacuum energy}

\author{G.E. Volovik
 $^{\#}$\/\thanks{%e-mail:
volovik@boojum.hut.fi}
}

\rauthor{G.E.Volovik}

\sodauthor{Volovik}

\address{Low Temperature Laboratory, Helsinki University of
Technology, P.O.Box 5100, FIN-02015, HUT, Finland
\\
 Landau Institute for Theoretical Physics RAS, Kosygina 2,
119334 Moscow, Russia}

\dates{September 14, 2009}{*}

\abstract{
The walls of the box which contains matter represent a membrane
 that allows the relativistic quantum vacuum to pass but not matter.
 That is why the pressure of matter in the box
 may be considered as the analog of the osmotic pressure.
However, we demonstrate that the osmotic pressure of matter
is modified due to interaction of matter with vacuum.
This interaction  induces the nonzero negative  vacuum pressure inside the box,
as a result the measured osmotic pressure becomes smaller than the matter pressure.
As distinct from the Casimir effect, this induced vacuum pressure
is the bulk effect and does not depend on the size of the box.
This effect dominates in the thermodynamic limit of the infinite volume of the box.
Analog of this effect has been observed in the dilute solution of $^3$He
in liquid $^4$He, where the superfluid $^4$He plays the role of
the non-relativistic quantum vacuum,
and $^3$He atoms play the role of matter.
}

\begin{document}

\maketitle

%\section{Introduction}

\section{Introduction}\label{sec:introduction}

In $q$-theory, the relativistic quantum  vacuum is considered as a self-sustained medium
\cite{KlinkhamerVolovik2008a,KlinkhamerVolovik2009c}.
This medium is described by the variable $q$, which is a conserved quantity analogous to particle density $n$ in condensed
matter, but as distinct from $n$ the vacuum `density' $q$ is the relativistic invariant quantity. The vacuum medium obeys the  thermodynamic Gibbs-Duhem relation
 $\epsilon_\mathrm{vac}(q) -\mu q=-P_\mathrm{vac}$, where $\epsilon_\mathrm{vac}(q)$ is the
 vacuum energy density and  $\mu$ is the vacuum  chemical potential -- the
 variable which is thermodynamically conjugate to $q$.
Dynamical equation for $q$  demonstrates that
$q$ gives rise to the cosmological term in  the Einstein equations of  general relativity with cosmological ``constant'' $\Lambda=\epsilon_\mathrm{vac} -\mu q=-P_\mathrm{vac}$ \cite{KlinkhamerVolovik2008b}.
The self sustained property of the quantum vacuum provides
a natural nullification
of the cosmological constant in equilibrium due to a self-adjustment mechanism:
 the vacuum variable is automatically self-tuned to nullify in equilibrium any contribution to the vacuum energy from different quantum fields. Dynamical equations also
demonstrate how the cosmological `constant' relaxes from  its original Planck scale value in the non-equilibrium vacuum  to the zero value in the final equilibrium state  \cite{KlinkhamerVolovik2008b}.  This provides the natural solution
 of the cosmological constant problem.

Till now we considered the homogeneous in space
  vacuum states, which were relevant for the homogeneous and isotropic Universe.
  Now we shall discuss the case, when the
  vacuum variable $q$  may vary in space. This occurs when matter
  (say, cold atomic gas) is confined in the box with non-penetrable walls.
 All the known walls which may confine matter, are however permeable for the vacuum.
 Thus the walls of the box which contains matter represent a semipermeable membrane
 that allows the vacuum (analog of solvent -- water) to pass but not matter (analog of solute). The
  pressure of matter in the box becomes equivalent to the osmotic pressure --  pressure that must be applied to a solution to prevent the inward flow of water.

In equilibrium, the chemical potential $\mu$ of the vacuum substance
must be the same inside and outside the wall, while the values of
  the vacuum variable $q$ are different because of interaction between vacuum and matter  inside the box.   As a result,  the total (osmotic) pressure of the gas inside the box is modified due to the vacuum: the negative vacuum pressure is added to the pressure of matter.

 This mechanism is of the thermodynamic origin and does not depend on whether the
 vacuum is relativistic or not.
 That is why it is also applicable to condensed matter systems, in particular
 to a dilute solution of $^3$He
 in superfluid $^4$He at zero temperature. In this system,
 superfluid $^4$He at $T=0$ plays the role of the non-relativistic quantum vacuum, and the
gas of the $^3$He quasiparticles plays the role of matter.
 The negative contribution of the `vacuum'  to the osmotic pressure of $^3$He is given by
  the same equation \eqref{excess} as for matter in the relativistic vacuum, but the vacuum compressibility \eqref{eq:Compressibility}
  introduced in \cite{KlinkhamerVolovik2008a} is substituted by the compressibility
 of liquid $^4$He. This  negative contribution to the osmotic pressure of $^3$He in liquid $^4$He has been experimentally measured.

\section{Correction to matter pressure due to vacuum}
\label{sec:correction}

\begin{figure}[top]
\centerline{\includegraphics[width=1.0\linewidth]{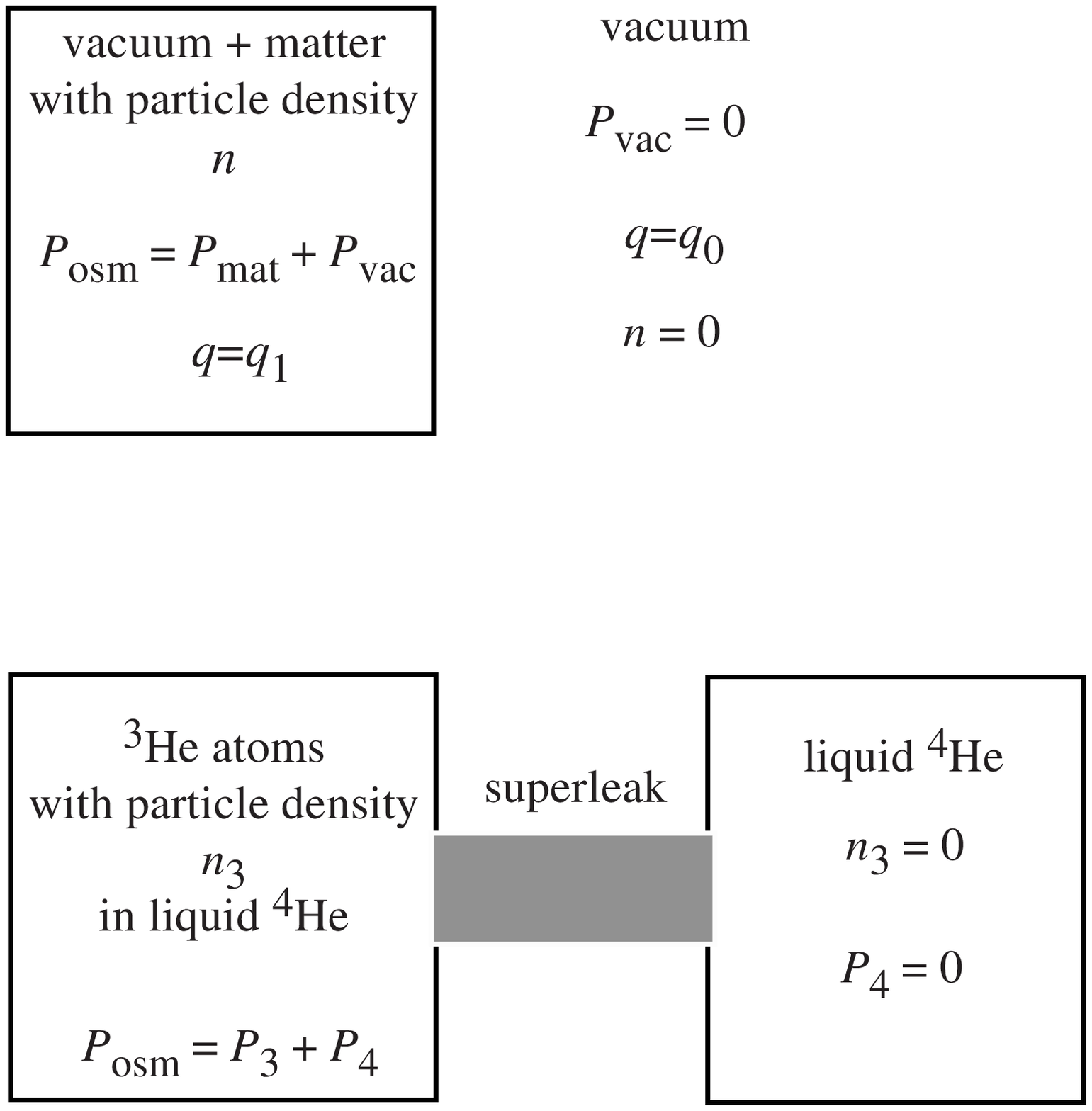}}
  \caption{\label{osm}
  Osmotic pressure of matter in the vacuum and its condensed matter analog.
  In both cases the negative contribution to osmotic pressure  is given by \eqref{excess}.
  \newline
  {\it Top}: matter inside the box. Vacuum may penetrate the walls of the box, and thus the
 vacuum  chemical potential $\mu$ is the same inside
 and outside the box. The pressure of matter inside the box (analog of osmotic pressure)
  is reduced, $P_\mathrm{osm} < P_\mathrm{mat}$, due to the negative contribution
  of the vacuum pressure, $P_\mathrm{vac}(q_1)<0$, which is  induced by interaction
  of vacuum with matter.
   \newline
   {\it Bottom}: dilute solution of $^3$He atoms in liquid $^4$He
   in the left box  is connected by superleak to the right box with pure liquid $^4$He.
   Superfluid $^4$He plays the role of the vacuum, which is disturbed
    by matter -- by $^3$He atoms -- inside the left box.
   Superfluid $^4$He may flow through the superleak, and thus
   its chemical potential $\mu_4$ is the same in two boxes.
   Osmotic pressure  of $^3$He inside the left box is reduced, $P_\mathrm{osm}< P_3$,
   due to the negative contribution $P_4<0$ of the background liquid $^4$He caused by
 interaction between liquid $^4$He and $^3$He atoms.
 }%%
\end{figure}

Let us consider the box, which contains matter with a particle density
$n$, say a cold gas (Fig. \ref{osm}  {\it top}).
The total energy density of matter and vacuum is
\begin{equation}
\epsilon(n,q)=\epsilon_\mathrm{vac}(q) + \epsilon_\mathrm{mat}(n,q)
 \,,
 \label{energy_density}
\end{equation}
where $\epsilon_\mathrm{vac}(q)$ is vacuum energy,
i.e. the energy density in the absence of matter:
\begin{equation}
 \epsilon_\mathrm{vac}(q) \equiv \epsilon(n=0,q)~~,~~
  \epsilon_\mathrm{mat}(n,q)\equiv   \epsilon(n,q)-
  \epsilon(n=0,q)
 \,,
 \label{energy_density2}
\end{equation}
and we take into account that the parameters of matter and thus the energy
density of matter $\epsilon_\mathrm{mat}$ depend on the vacuum state
and vacuum variable $q$.
The vacuum pressure is determined by Gibbs-Duhem thermodynamic relation
\cite{KlinkhamerVolovik2008a}:
\begin{equation}
P_\mathrm{vac}(q)=- \tilde\epsilon_\mathrm{vac}(q)=-\epsilon_\mathrm{vac}(q)+ \mu q
 \,,
 \label{vac_energy}
\end{equation}
where $\mu$ is the chemical potential in thermodynamics, and the integration constant in dynamics
 \cite{KlinkhamerVolovik2008b};
the thermodynamic potential
$\tilde\epsilon_\mathrm{vac}(q)=\epsilon_\mathrm{vac}(q)- \mu q$ enters Einstein equations as cosmological constant,
$ \Lambda\equiv  \tilde\epsilon_\mathrm{vac}(q)$  \cite{KlinkhamerVolovik2008b}.
Outside of the box, where matter is absent, the value of $q=q_0$
in the equilibrium self-sustained vacuum is determined by the zero pressure  condition:
\begin{equation}
P_\mathrm{external} =-\epsilon_\mathrm{vac}(q_0) +
q_0\frac{d\epsilon_\mathrm{vac}(q)}{dq}\Big|_{q_0}=0
 \,.
 \label{external_pressure}
\end{equation}
Since vacuum penetrates the walls of the box, its chemical potential
$\mu=\partial \epsilon/\partial q|_n$ must be the same across the wall.
This gives the condition:
\begin{equation}
\mu={\rm constant}=\frac{d\epsilon_\mathrm{vac}(q)}{dq}\Big|_{q_0}=
\frac{d\epsilon_\mathrm{vac}(q)}{dq}\Big|_{q_1}
+ \frac{\partial\epsilon_\mathrm{mat}(n,q)}{\partial q}\Big|_{q_1}
 \,,
 \label{mu_condition}
\end{equation}
where $q_1$ is the equilibrium value of  $q$  inside the box, which
is determined by \eqref{mu_condition}. For small deviations $|q_1-q_0|\ll q_0$,
equation  \eqref{mu_condition} gives:
\begin{equation}
(q_1-q_0)\frac{d^2\epsilon_\mathrm{vac}(q)}{dq^2}\Big|_{q_0}=
- \frac{\partial\epsilon_\mathrm{mat}(n,q)}{\partial q}\Big|_{q_0}
 \,.
 \label{q_1-q_0}
\end{equation}

Let us find the pressure inside the box. The conventional pressure of matter is
\begin{equation}
P_\mathrm{mat} =-\epsilon_\mathrm{mat}(n,q_1) + \mu_\mathrm{mat} n
=-\epsilon_\mathrm{mat}(n,q_1) +n \frac{\partial\epsilon_\mathrm{mat}(n,q_1)}{\partial n}
 \,.
 \label{matter_pressure}
\end{equation}
However, the pressure inside the box,
 which is the analog of the osmotic pressure of matter, differs from matter pressure
 due to the modified vacuum pressure. The total pressure inside the box is
\begin{equation}
P_\mathrm{osmotic} =-\epsilon(n,q_1) + \mu q_1
+ \mu_\mathrm{mat} n=P_\mathrm{vac}(q_1)+P_\mathrm{mat}   \,.
 \label{osmotic_pressure}
\end{equation}
Expanding $P_\mathrm{vac}(q_1)$ in $q_1-q_0$ one obtains:
\begin{eqnarray}
P_\mathrm{vac}(q_1) = \mu q_1-\epsilon_\mathrm{vac}(q_1)
\nonumber
\\
 \approx  \mu q_0-\epsilon_\mathrm{vac}(q_0) +(q_1-q_0)\left( \mu
 -  \frac{d\epsilon_\mathrm{vac}(q)}{dq}\Big|_{q_0} \right)
 \nonumber
 \\
- \frac{1}{2}(q_1-q_0)^2\frac{d^2\epsilon_\mathrm{vac}(q)}{dq^2}\Big|_{q_0}
 \nonumber
 \\
=  - \frac{1}{2}(q_1-q_0)^2\frac{d^2\epsilon_\mathrm{vac}(q)}{dq^2} \Big|_{q_0}
\,,
 \label{osmotic_pressure2}
\end{eqnarray}
where we used  \eqref{vac_energy}, \eqref{external_pressure} and \eqref{mu_condition}.
Using  \eqref{q_1-q_0} for $q_1-q_0$ one obtains the following
expression for the negative contribution to the osmotic pressure
due to the induced vacuum pressure:
\begin{equation}
P_\mathrm{osmotic} =P_\mathrm{mat} + P_\mathrm{vac}
=P_\mathrm{mat}-
\frac{1}{2}\chi_\mathrm{vac}  \left[q\frac{\partial\epsilon_\mathrm{mat}(n,q)}{\partial q}  \right]_{q=q_0}^2   \,,
 \label{excess}
\end{equation}
where $\chi_\mathrm{vac}$ is vacuum compressibility \cite{KlinkhamerVolovik2008a}:
\begin{equation}
\chi_\mathrm{vac}^{-1} =
\left[q^2\;\frac{d^2\epsilon_\mathrm{vac}(q)}{dq^2}\,\right]_{q=q_0}
\geq 0~.
\label{eq:Compressibility}
\end{equation}

\section{Mapping to dilute solution of  $^3$He in superfluid $^4$He at $T=0$}
\label{sec:mapping}

There is the following correspondence with the dilute solution of
$^3$He in superfluid $^4$He at $T=0$.
Superfluid $^4$He represents the vacuum, and $^3$He
quasiparticles represent matter living in the background of
the superfluid $^4$He  `vacuum'.
Thermal phonons which also represent  `matter'  are absent at $T=0$.
In typical experimental situation, the box which
contains mixture is connected with pure liquid $^4$He by the Vicor glass superleak,
which is not penetrable by $^3$He atoms, but superfluid $^4$He
may flow through the superleak (Fig. \ref{osm} {\it bottom}).

The role of the vacuum variable $q$ is played by the particle density
of $^4$He atoms in pure superfluid $^4$He and in mixture, $q\equiv n_4$.
 The vacuum chemical potential $\mu$ is equivalent to  the chemical potential of $^4$He,
 $\mu\equiv \mu_4$. It is the same in pure $^4$He and in mixture, since superfluid $^4$He
 may flow through the superleak, $ \mu_4={\rm constant}$. The vacuum energy density corresponds
 the energy density of pure liquid $^4$He,
 $\epsilon_\mathrm{vac}(q)\equiv \epsilon(n_3=0,n_4)$.
The helium liquid obeys the same equation of state
as relativistic quantum vacuum:  $\tilde\epsilon =-P$, where
 $\tilde\epsilon = \epsilon - \mu_4 n_4$. This is the consequence of thermodynamic Gibbs-Duhem relation  at $T=0$,
 which is valid for any system, relativistic and non-relativistic.

The matter density is played by the particle density of $^3$He atoms in mixture,
$n\equiv n_3$. The energy density of matter is determined as correction to the energy density
of the `vacuum'
 when the $^3$He atoms with density  $n_3$ are added to liquid $^4$He:
$\epsilon_\mathrm{mat}(n,q)\equiv \epsilon(n_3,n_4)-\epsilon(n_3=0,n_4)$.
The matter pressure $P_\mathrm{mat}\equiv P_\mathrm{quasiparticles}$
is determined as the pressure of the Fermi liquid (non-ideal Fermi gas)
with the same parameters as the non-ideal Fermi gas of $^3$He quasiparticles,
i.e. with the same density, effective mass and  the other Fermi liquid parameters.
These parameters depend on the `vacuum variable' $n_4$, which in particular enters the energy spectrum of fermionic $^3$He quasiparticles  \cite{LandauPomeranchuk1948,BardeenBaymPines1967}
\begin{equation}
E({\bf p},n_4)= E_3(n_4)+\frac{p^2}{2m^*(n_4)} + \ldots
 \label{spectrum}
\end{equation}

It is assumed that the pure $^4$He outside the superleak has zero pressure,
though this assumption is not very important for measuring the osmotic pressure
which is the difference between the inside and outside pressures.
If the pressure of pure liquid $^4$He outside is zero, then the pressure in the mixture
$P_\mathrm{internal} =P_\mathrm{osmotic}$ is osmotic pressure of $^3$He.
Using the above correspondence, the correction to the osmotic pressure due to the modified pressure
of superfluid $^4$He in the mixture can be found from \eqref{excess}, where the vacuum compressibility
must be substituted by compressibility of pure superfluid $^4$He, $\chi_\mathrm{vac}\equiv \chi_4=n_4^2 d\mu_4/dn_4$.

In the limit of small concentration, $n_3\rightarrow 0$, the main contribution
to the induced `vacuum pressure' in the rhs of \eqref{excess} comes from the `vacuum'
dependence of the parameter $E_3(n_4)$ in the quasiparticle spectrum \eqref{spectrum}.
 Since in the dilute limit one has $\mu_3(n_4)\approx E_3(n_4)$, one obtains
 $\partial\epsilon_\mathrm{mat}(n_3,n_4)/\partial n_4 \approx   n_3 dE_3/dn_4\approx  n_3 d\mu_3/d n_4$.
 Then equation \eqref{excess} gives the following correction
  to the osmotic pressure at small concentrations of $^3$He
\begin{equation}
P_\mathrm{osmotic} \approx P_\mathrm{quasiparticles}
 -\frac{n_3^2 }{2}  \frac{\left( d\mu_3/dn_4  \right)^2} {d\mu_4/dn_4 } ~~,~~n_3\rightarrow 0  \,.
 \label{excess_mixture}
\end{equation}
The reduction of the osmotic pressure in dilute solutions of $^3$He in superfluid $^4$He due to the modification of the superfluid background has been experimentally observed
\cite{EbnerEdwards1971,Owers-Bradley1997,Rysti2008}.

\section{Discussion}
\label{sec:discussion}

We discussed the phenomenon similar to the Casimir effect,
in which the walls also induce the vacuum pressure.
However, as distinct from the original Casimir effect where
 the vacuum pressure depends on the dimension $L$ of the box,
 $P_\mathrm{Casimir}Ê\propto L^{-4}$, the vacuum pressure induced
  by matter is the bulk effect and does not depend on the size $L$ of the box.
  This effect  results from the interaction between the vacuum and matter
 and it  dominates in the thermodynamic limit of the large volume of the box.
The interaction between the vacuum and matter occurs
 in particular due to the dependence of the parameters
of Standard Model matter on the vacuum variable $q$, for example
via the ultra-violet cut-off which enters the running coupling constants.

In the present Universe this effect is extremely small. Since the present atomic
 matter is very dilute compared to the vacuum,  the interaction between matter and vacuum produces only small perturbation of the vacuum state. The estimation gives
$P_\mathrm{vac}\sim - \chi_\mathrm{vac} \epsilon_\mathrm{mat}^2$,
with   $\chi_\mathrm{vac}\sim  E_\mathrm{Planck}^{-4}$
if the vacuum variable has the Planck energy scale, or
$\chi_\mathrm{vac}\sim E_\mathrm{QCD}^{-4}$ in case of gluon condensate
in quantum chromodynamics as a soft component of the vacuum substance
 with the characteristic QCD scale $E_\mathrm{QCD}$ \cite{KlinkhamerVolovik2009b}. However, the modified pressure of gluon condensate
may be considerable inside the neutron stars, where $q d\epsilon_\mathrm{mat}/dq$
and $1/\chi_\mathrm{vac} $ are both determined by the QCD scale and may
have the comparable magnitudes, $q d\epsilon_\mathrm{mat}/dq \sim 1/\chi_\mathrm{vac} \sim
E_\mathrm{QCD}^4$. It may also influence the models for the interior of  black holes and black-hole candidates
\cite{MazurMottola2004,VisserLiberatiSonego2009}. In principle the vacuum density $q$ may vanish
in the center of the black hole singularity.

The vacuum effect on the order of
$\epsilon_\mathrm{mat}^2/E_\mathrm{Planck}^4$ appears also in the
nonequilibrium Universe, see e.g. \cite{ZS}. Such effects were
essential in the early Universe, and if for some reasons the vacuum
energy was frozen at later evolution, this could give the reasonable
estimate for the present magnitude of the vacuum energy and
cosmological constant $\Lambda$. The dependence of the matter energy
density on the vacuum variable $q$ which enters the induced vacuum
pressure in \eqref{excess}, should appear at
 the temperature of the electroweak crossover $T_\mathrm{ew}$,
 where the emerging masses of elementary particles depend on $q$.
 The vacuum pressure estimated at this temperature is
$P_\mathrm{vac}\sim -\epsilon_\mathrm{mat}^2/E_\mathrm{Planck}^4
\sim -T_\mathrm{ew}^8/E_\mathrm{Planck}^4$. Similar results but from
a different argumentation were obtained  in Ref.
\cite{KlinkhamerVolovik2009a}, where it was demonstrated that the
electroweak crossover necessarily generates the vacuum energy
density $\tilde\epsilon_\mathrm{vac}=\Lambda \sim
T_\mathrm{ew}^8/E_\mathrm{Planck}^4$. This  vacuum energy  is
comparable with the present value of the cosmological constant
$\Lambda$. If the freezing mechanism for the vacuum energy suggested
in Ref. \cite{KlinkhamerVolovik2009a} is confirmed, this will
support the theories where the dark energy is related to the
electroweak physics, such as in Ref.  \cite{ArkaniHamed-etal2000}.

The analogous reduction of the osmotic pressure  has been experimentally observed in
the dilute solution of $^3$He
in superfluid $^4$He, where superfluid $^4$He at $T=0$ plays the role of quantum vacuum,
and $^3$He atoms play the role of matter.
The observed negative correction to the osmotic pressure is usually described in terms of the additional
 effective interaction between the $^3$He fermionic quasiparticles,
which is mediated by the
 background superfluid $^4$He, e.g.
by an exchange of the virtual $^4$He excitations -- phonons
 \cite{Saam1969,BardeenBaymPines1967,ViveritPethickSmith2000,Albus2002}.
 However, in the considered case of matter in the background
 of relativistic quantum vacuum, introduction of the additional
 interaction between the matter fields for the description of the effect seems unreasonable. First, for the particular choice of the $q$-field in terms of the 4-form field
 \cite{KlinkhamerVolovik2008a,KlinkhamerVolovik2008b},
 there is no propagating excitations of the $q$-field which can mediate the interaction,
 but the effect takes place.   Second, the introduced interactions will be different for different species of matter and may be even non-local.

It is more physical to describe the negative
 contribution to matter pressure in the general framework of the response
 of  the quantum vacuum to perturbations. In the same manner
 the Casimir effect both in quantum vacuum and in condensed matter systems
 \cite{Krech1994,Volovik2003} is better described in terms of the properties of quantum vacuum rather than in terms of van der Waals or other forces between the matter fields.    The Casimir effect,
when the walls perturb the quantum and thermal fluctuations, and the reduction of the matter pressure are just two different types of the response of the quantum vacuum. The other types of the perturbation of the quantum vacuum and the vacuum response to the perturbations are discussed in Refs. \cite{Volovik2003,KlinkhamerVolovik2008a}.

\section*{\hspace*{-4.5mm}ACKNOWLEDGMENTS}
\noindent
It is a pleasure to thank  Frans Klinkhamer and Alexander Sebedash for helpful discussions.
GEV is supported in part by Academy of Finland, Centers of excellence program 2006-2011,
the Russian Foundation for
Basic Research (grant 06--02--16002--a) and the Khalatnikov--Starobinsky
leading scientific school (grant 4899.2008.2).
The research leading to these results has received funding from the EUÕs Seventh Framework Programme (FP7/2007-2013) under grant agreement  \# 228464 (MICROKELVIN).

%\newpage%tmp

\end{document}